\documentclass[twocolumn,prd,aps,preprintnumbers, showpacs, nofootinbib,superscriptaddress, notitlepage, 10pt]{revtex4-1}

\usepackage{enumitem}

\usepackage{graphicx}
\usepackage{dcolumn}
\usepackage{bm}
\usepackage{url}
\usepackage{epsfig}
\usepackage{amsmath}
\input{epsf}
\usepackage{psfrag}
\usepackage{xcolor}
\usepackage{pdfpages}
\usepackage{hyperref}
\usepackage{enumitem}

\usepackage{graphicx}
\usepackage{dcolumn}
\usepackage{bm}
\usepackage{url}
\usepackage{epsfig}
\usepackage{amsmath}
\input{epsf}
\usepackage{psfrag}
\usepackage{xcolor}
\usepackage{pdfpages}
\usepackage[normalem]{ulem}
\makeatletter
\AtBeginDocument{\let\LS@rot\@undefined}
\makeatother
\newcommand{\bea}{\begin{eqnarray}}
\newcommand{\eea}{\end{eqnarray}}

\newcommand{\nn}{\nonumber}

\begin{document}

\title{Particle Correlations in Jets}

\author{Wenbin Zhao}
\email{WenbinZhao@lbl.gov}
\affiliation{Institute of Particle Physics and Key Laboratory of Quark and Lepton Physics (MOE), Central China Normal University, Wuhan, 430079, Hubei, China}
\affiliation{Nuclear Science Division, Lawrence Berkeley National
Laboratory, Berkeley, CA 94720, USA} 
\affiliation{Physics Department, University of California, Berkeley, California 94720, USA}

\author{Volker Koch}
\email{vkoch@lbl.gov}
\affiliation{Nuclear Science Division, Lawrence Berkeley National
Laboratory, Berkeley, CA 94720, USA} 

\author{Feng Yuan}
\email{fyuan@lbl.gov}
\affiliation{Nuclear Science Division, Lawrence Berkeley National
Laboratory, Berkeley, CA 94720, USA} 

\begin{abstract}
We study particle correlations in high energy jets by comparing the measured energy-energy correlator (EEC) with that constructed from two individual energy flows with respect to the jet axis. This comparison demonstrates that the genuine correlations are stronger at moderate/large angle across all jet energies, indicating that they are coming from correlated splitting. Meanwhile, at small angle, they increase with jet energy, exhibiting a peculiar pattern that can be used to explore jet energy loss in heavy ion collisions.  This method will provide a unique tool to disentangle different physics, by comparing the genuine correlations in jet EEC between heavy-ion collisions and proton-proton collisions. 
\end{abstract}

\maketitle

\textbf{\textit{Introduction.}} 
Energy-energy correlators (EECs) in $e^+e^-$ annihilation were introduced to test the asymptotic freedom of Quantum Chromodynamics (QCD) more than three decades ago~\cite{Basham:1978bw,Basham:1977iq,Basham:1978zq}. Theoretical developments have advanced its applications to extract various physics and study precision strong interaction physics~\cite{Collins:1981uk,Ali:1982ub,Clay:1995sd,deFlorian:2004mp,DelDuca:2016csb,Tulipant:2017ybb,Kardos:2018kqj,Moult:2018jzp,Dixon:2018qgp,Dixon:2019uzg,Ebert:2020sfi,Schindler:2023cww}. In recent years, a renaissance of interest in EEC has spanned across different fields~\cite{Lee:2006nr,Berger:2003iw,Hofman:2008ar,Chen:2020vvp,Lee:2022ige,Craft:2022kdo,Komiske:2022enw,Andres:2022ovj,Andres:2023xwr,Andres:2023ymw,Yang:2023dwc,Andres:2024ksi,Barata:2023bhh,Barata:2023zqg,Bossi:2024qho,Lee:2023tkr,Lee:2023xzv,Barata:2024ukm,Lee:2024esz,Chen:2024nyc,Holguin:2023bjf,Holguin:2024tkz,Xiao:2024rol,Xing:2024yrb,Andres:2024hdd,Liu:2024lxy,Alipour-fard:2024szj,Kang:2024dja,Barata:2024wsu,Csaki:2024zig,Fu:2024pic,Apolinario:2025vtx,Barata:2025fzd,Chen:2025rjc,Lee:2025okn,Guo:2025zwb,Chang:2025kgq,Kang:2025zto,Herrmann:2025fqy,Cuerpo:2025zde}, see, a recent review~\cite{Moult:2025nhu}. In particular, it has been extensively applied to systematic measurements of jet substructure in proton-proton collisions and heavy-ion collisions at the LHC and RHIC~\cite{CMS:2024mlf,ALICE:2024dfl,Tamis:2023guc,CMS:2025ydi,ALICE:2025igw,ALICEpA}. In the following, we study this physics from a different perspective, focusing on the correlation between particles inside the high energy jets. The goal is to help us to understand the underlying physics and to reveal hadronization effects in these correlation measurements. 
 
The EECs inside the jet is defined as 
\begin{eqnarray}
\label{eq:eecdef} 
&&\frac{d
\langle {\rm EEC}(\theta)\rangle_{\rm full} 
}{d^2{\theta}}    \\
&&=   \frac{1}{N_{jet}}\sum_{jets\, {\rm J}}
\sum_{i\neq j\in {\rm J}} 
\frac{E_iE_j}{E_J^2}  \delta^{(2)}(\vec{\theta}-(\vec{\theta}_i-\vec{\theta}_j)) \ ,\nn  
\end{eqnarray}
where $N_{jet}$ represents number of jets. 
The sum is over all particles $(i,j)$ in the jet so that ${\rm EEC}(\theta)$ corresponds to the energy weighted sum (distribution) of all pairs with relative angle $\theta$, and jet energy $E_J=\sum_{i}E_i$.
At colliders, the definition of the angle includes the rapidity difference between the two particles, e.g., $R_L=\sqrt{\delta \eta^2+\delta \phi^2}$ is used in $pp$ collisions where $\delta \eta$ represents the rapidity difference and $\delta \phi$ for the azimuthal angle difference. However for small angle $R_L = \theta$ to very good approximation.  In the transverse plane perpendicular to the jet axis, each angle represents the distance between the particle and the jet axis. Because they are all within the jet with a small jet size, the angular distribution from the above will peak at small angle and the dominant contributions from perturbative QCD come from collinear radiation. The soft gluon radiation is suppressed due to the energy weight. Therefore, a perturbative QCD prediction can be applied to the EEC observables~\cite{Basham:1978bw}. 

In the perturbative collinear region ($\theta >0.1$), additional gluon radiations to all orders can be resummed through a factorization approach~\cite{Dixon:2019uzg} with logarithmic corrections. With additional contributions from renormalons~\cite{Schindler:2023cww} and power corrections~\cite{Chen:2024nyc,Lee:2024esz}, these precision improvements play an important role to describe the experimental data from EECs in $e^+e^-$ annihilation. Toward the other end in the angular distribution ($\theta\sim 0$), additional physics has to be taken into account such as the assumption of a ``free hadron" phase~\cite{Komiske:2022enw}. Recently, a non-perturbative transverse momentum dependent (TMD) fragmentation mechanism has been proposed to describe the distribution in this region~\cite{Liu:2024lxy,Barata:2024wsu} and further theoretical developments along this direction have been reported~\cite{Chang:2025kgq,Lee:2025okn,Guo:2025zwb,Kang:2025zto,Herrmann:2025fqy,Cuerpo:2025zde}.

While the EEC, Eq. \eqref{eq:eecdef}, is commonly called a ``correlator" it is actually an energy weighted pair distribution, and thus will not vanish even in the absence of any particle correlations in the jets. To extract the true, genuine correlations one typically subtracts the product of the single particle distributions, which remove all uncorrelated pairs. Alternatively one uses mixed events to determine the uncorrelated pair distribution. In the context of hadronic physics this is for example done for particle interferometry measurements to determine the size of the particle emitting source~\cite{Wiedemann:1999qn}. 
In this paper we will take exactly this approach in order to extract and identify the true two-particle correlations.  

Performing such an analysis will enable us to highlight the physics behind the three different regions in the EEC in jet. In particular, we find that the correlation is significant in both small and moderate $\theta$ regions. Although their physics are totally different --  non-perturbative TMD fragmentation dominates small $\theta$ while collinear splitting dominates the moderate $\theta>0.1$ -- they both come from correlations, i.e., two particles originating from one source. In this study, we employ the PYTHIA8.306 event generator \cite{Sjostrand:2019zhc} to demonstrate our proposed approach. 
We further perform a proof-of-principle analysis using the CMS 2011 Open Data~\cite{CMS2011:Open_data} and show qualitative agreement with our PYTHIA8 simulation baseline.

\textbf{\textit{Disconnected two particle distributions from single particle distributions.}}
The disconnected two particle distributions  based on the single particle distributions need to be carefully defined. For this purpose, the jet axis can be used as an appropriate reference. 
In the experimental measurements, the energy correlations are counted against the particles inside a jet. Therefore, a jet axis can be identified. Because of the rotation symmetry around the jet axis, the above angular correlation is a two-dimensional distribution,
\begin{equation}
    \vec{\theta}=\vec{\theta}_i-\vec{\theta}_j  \ ,
\end{equation}
where $\vec{\theta}_{i}=\vec{P}^{h}_{i\perp}/(z_{i}E_J)$ represents the angle of hadron $i$ with respect to the jet axis, and $z_{i}$ is its momentum fraction of the jet energy.  We consider a jet radius of $R\ll 1$. Therefore, $|\vec{P}^h_{i\perp}|\ll E_i=z_{i}E_J$, i.e., the transverse momentum of the hadron $|\vec{P}^h_{i\perp}|$ is much smaller than its longitudinal momentum $E_{i}$. 

With the jet axis well defined, one can also define the correlator depending on two angles,
\begin{eqnarray}
\frac{d
\langle {\rm EEC}(\theta_1,\theta_2)\rangle 
}{d^2{\theta_1}d^2{\theta_2}}
&=&    \frac{1}{N_{jet}}\sum_{jets \,{\rm J} }\sum_{i\neq j\in {\rm J}} \frac{E_iE_j}{E_J^2}\nn\\
&&\times \delta^{(2)}({\theta_1}-\theta_i) \delta^{(2)}({\theta_2}-\theta_j) \ ,\label{eq:eecdef0}
\end{eqnarray}
from which we can derive the angular distribution of EEC of Eq.~(\ref{eq:eecdef})~\cite{Basham:1978zq},
\begin{equation}
\frac{d
\langle {\rm EEC}(\theta)\rangle 
}{d^2{\theta}}
= \int d^2\theta_1d^2\theta_2\frac{d
\langle {\rm EEC}(\theta_1,\theta_2)\rangle 
}{d^2{\theta_1}d^2{\theta_2}}  \delta^{(2)}(\vec{\theta}-(\vec{\theta}_1-\vec{\theta}_2)) \ .\label{eq:eecdef1}
\end{equation}
However, Eq.~(\ref{eq:eecdef0}) contains disconnected un-correlated contribution from two energy flows through a  construction
\begin{equation}
    \frac{d
\langle {\rm EEC}(\theta_1,\theta_2)\rangle 
}{d^2{\theta_1}d^2{\theta_2}}|_{\rm Disc.}\propto \frac{d
\langle {\rm E}(\theta_1)\rangle}{d^2{\theta_1}}\times \frac{d\langle{\rm E}(\theta_2)\rangle 
}{d^2{\theta_2}}\ ,
\end{equation}
where the single energy distribution is defined as, 
\begin{eqnarray}
\frac{d\langle {\rm E}(\theta)\rangle }{d^2{\theta}}&=&    \frac{1}{N_{jet}}\sum_{jets \,{\rm J} }\sum_{i\in {\rm J}}
\frac{E_i}{E_J} \delta^{(2)}(\vec{\theta}-\vec{\theta}_i) \ .\label{eq:ecdef}
\end{eqnarray}
The above equations will be used to construct the subtraction term for the genuine correlations. 

To construct the disconnected correlator, one has to properly take into account the energy weighting and, related to that,  the normalization of the angular integral of Eq.~(\ref{eq:eecdef}). Clearly, the latter is not normalized to 1.
To account for this difference, we introduce a normalization factor $N_{\rm Disc.}$ for the disconnected correlation
\begin{eqnarray}
&&   \frac{d}{d^2\theta} \langle {\rm EEC}(\theta)\rangle_{\rm Disc.} =N_{\rm Disc.}\int d^2\theta_1 d^2\theta_2\nn\\
    &&~~~~~\times 
    \frac{d \langle  {\rm E}(\theta_1)\rangle}{d^2\theta_1}
    \frac{d\langle  {\rm E}(\theta_2)\rangle}{d^2\theta_2} \delta^{(2)}(\vec{\theta}-(\vec{\theta}_1-\vec{\theta}_2)) \ .\label{eq:eec0p}
\end{eqnarray}
Without $N_{\rm Disc.}$, the right hand of the above equation will be normalized to 1 by definition. For the typical jet energy at the LHC, we find that $N_{\rm Disc.}\approx 0.85$, which is equal to the integral of Eq.~(\ref{eq:eecdef}) over $\theta$ for a particular jet energy range and also related to the integrated di-hadron fragmentation functions, see, Ref.~\cite{Guo:2025zwb}.

In experiment, one can also construct the above disconnected correlation by means of mixing event or rather jet mixing. As detailed in Appendix A, one generates ``mixed" jets by picking particles from random jets. For example, given a {\em true} jet with $n$ particles the corresponding {\em mixed} jet is made with $n$ particles with one each from $n$ randomly selected true jets. Given the mixed jets, the disconnected, uncorrelated EEC correlator is then obtained from Eq.~\eqref{eq:eecdef} where the true jets are replaced by the mixed jets with an additional normalization factor,
    \begin{eqnarray}
    &&\frac{d\langle {\rm EEC}(\theta)\rangle_{\rm mix} }{d^2{\theta}} = {\cal C}_{mix}\frac{1}{N_{jet}}\sum_{\rm mixed\, jets\, J }\nn\\
    &&~~~~ \times    \sum_{i\neq j\in {\rm J} } \frac{E_{i}}{E_{J,i}}\times \frac{E_{j}}{E_{J,j}} \delta^{(2)}({\vec{\theta}}-(\vec{\theta}_i-\vec{\theta}_j))\ .\label{eq:eecmix}
\end{eqnarray}
To account for the proper summation of mixed jets at different rapidities, here we keep the original energy fraction for each particle, e.g., $E_i/E_{J,i}$ for particle $i$ from Jet$_{i}$.

In the above equation, we introduce a normalization factor ${\cal C}_{mix}$ to account for the differences between the mixing jet events and the true jet events. The reason behind this is that the above sum favors particles with larger momentum fractions in the mixed jet, although the angular distribution remains.
We have verified that the above two equations (\ref{eq:eec0p}) and (\ref{eq:eecmix}) lead to the same shape of the angular distribution of the disconnected EEC for the jet production at the LHC (see Fig. 3 in the Appendix). Of course, the application of the mixed-event method to real data must consider detector inefficiencies, limited acceptance, and other systematic effects that could obscure small correlation signals. Although our study focuses on the theoretical framework under idealized conditions, the agreement between the disconnected two-particle correlations obtained from both the mixed-event and single-particle approaches provides a valuable cross-check. This comparison can help experimental groups isolate the genuine physics signals once the detector-related systematics are properly addressed.

Similarly we can evaluate the single energy flow result from Eq.~\eqref{eq:ecdef} by using mixed jets. As we show in the Appendix they agree with those obtained from the true jets, as they should.

Meanwhile, it has been shown in the literature that the particle distributions inside a jet can be described by the parton fragmentation functions in both the collinear case~\cite{Kaufmann:2015hma,Chien:2015ctp,Gao:2024dbv} and the TMD case~\cite{Neill:2016vbi,Kang:2017glf,Kang:2019ahe}. In particular, a recent paper~\cite{Mi:2025abd} has shown that the TMD fragmentation functions can be applied to the angular distribution of Eq.~(\ref{eq:ecdef}) with a typical parameterization~\cite{Sun:2014dqm}, see, also Ref.~\cite{Gao:2025evv}. In the Appendix, we show that these TMD fragmentation functions can also give a reasonable description of the disconnected EEC at small angle.

\begin{figure}[htbp]
  \begin{center}
   \includegraphics[scale=0.35]{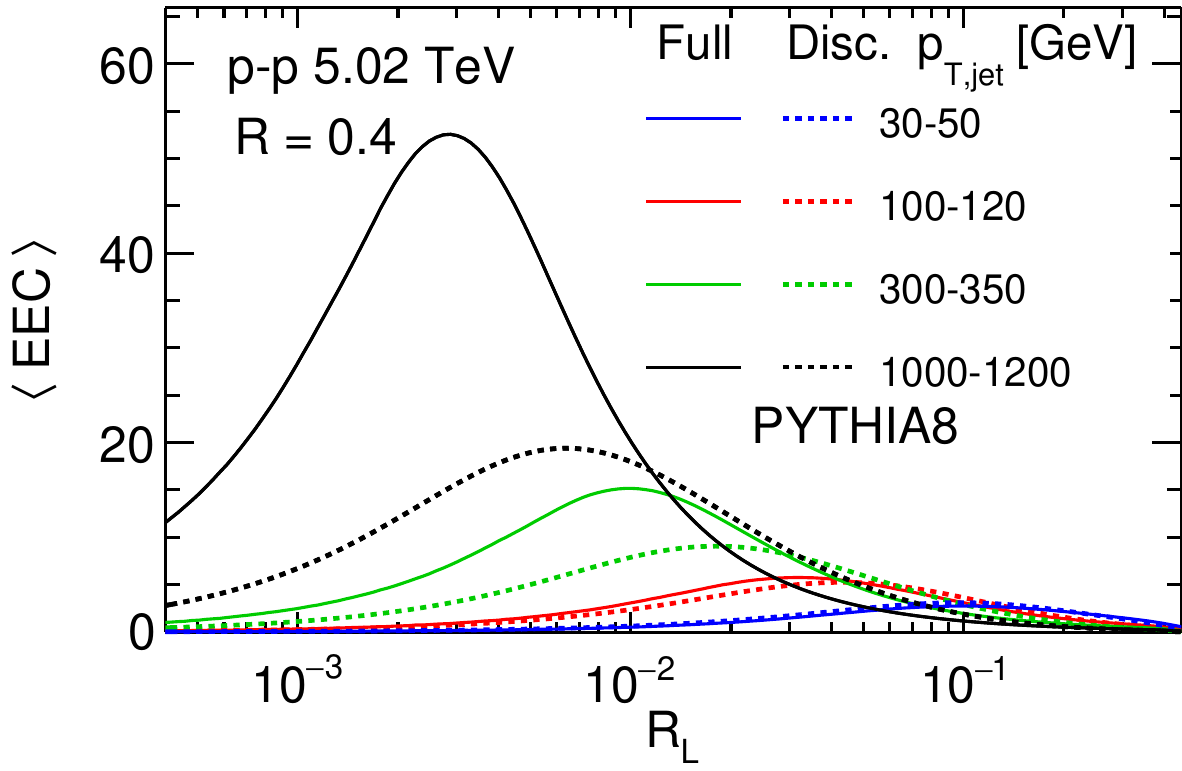} 
\caption{The PYTHIA simulations of angular distributions for the EECs in jet for different jet energies at the LHC. Here, we compare the full correlation, $\left< \rm EEC\right>_{\rm full}$, corresponding to the experiment measurements of Eq.~(\ref{eq:eecdef}) and the disconnected correlation, $\left<\rm EEC\right>_{\rm Disc.}$, constructed from two separate energy flows of Eq.~(\ref{eq:eec0p}). Here, $R_L=\theta$ in these equations. }
  \label{fig:eeccompare}
 \end{center}
  \vspace{-5.ex}
\end{figure}

\textbf{\textit{Two Particle Correlations in a Jet.}} With these results, we now compare the correlations between Eq.~(\ref{eq:eecdef}) and (\ref{eq:eec0p}). As we see from Fig.~\ref{fig:eeccompare}, they have a similar shape, peaked at small angles, but there are obvious differences which are due to the actual true correlations. 

\begin{figure}[htbp]
   \includegraphics[scale=0.35]{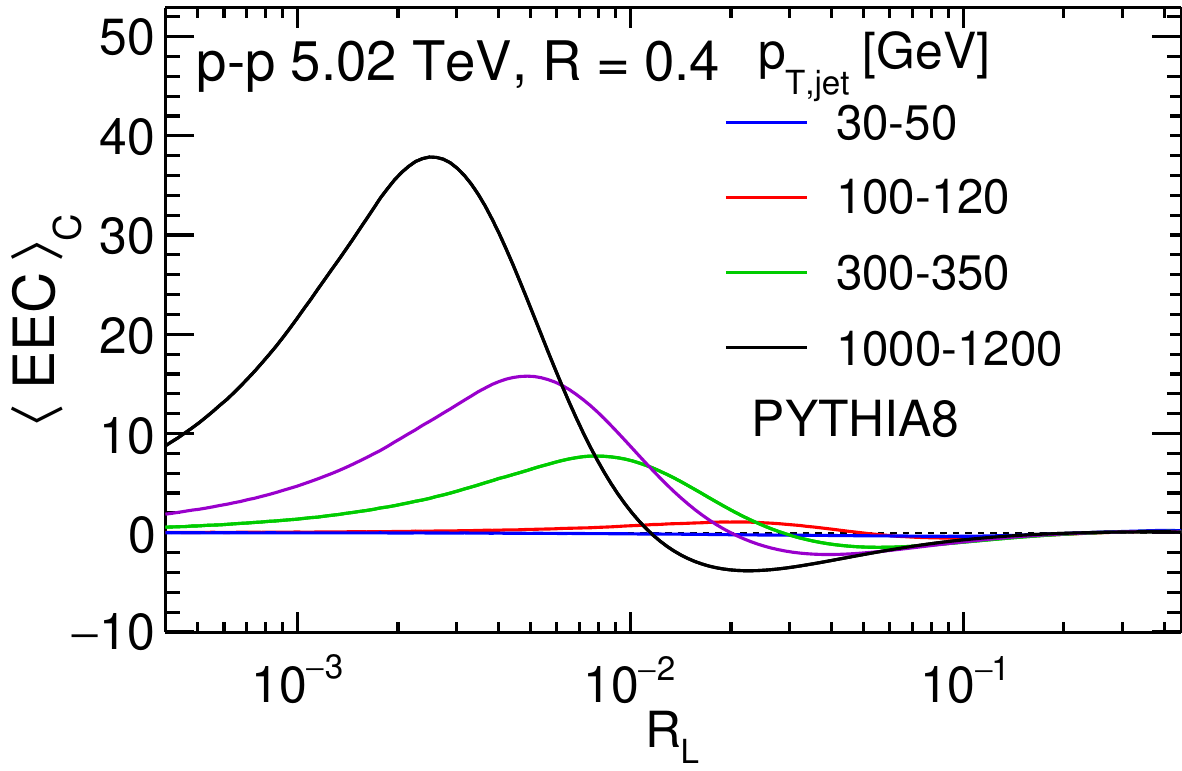} 
      \includegraphics[scale=0.35]{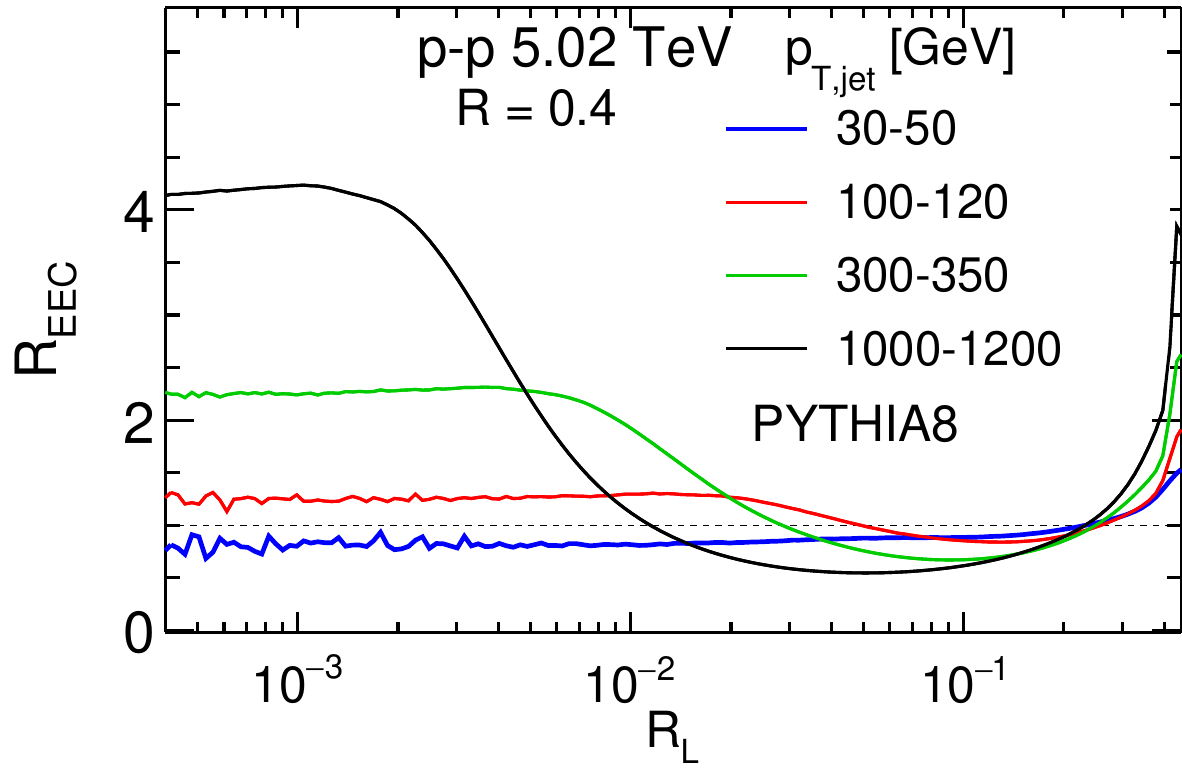} 
\caption{The two particle genuine correlations in jets as functions of the angle between them for different jet energies at the LHC: (upper) defined as difference, Eq.~(\ref{eq:eecc}) and (lower) defined as ratio, Eq.~(\ref{eq:eecc0}). Here, as above, $R_L=\theta$. 
}
  \label{fig:eecratio}
  \vspace{-3.ex}
\end{figure}

These genuine correlations can be extracted by the difference between the two correlators,
\begin{eqnarray}
&&    \langle {\rm EEC}(\theta)\rangle_C =\frac{d
\langle {\rm EEC}(\theta)\rangle_{\rm full} 
}{d^2{\theta}}-\frac{d
\langle {\rm EEC}(\theta)\rangle_{\rm Disc.} 
}{d^2{\theta}}
    \ .\label{eq:eecc}
\end{eqnarray}
Again, the second term in the right hand side of the above equation can be obtained by the integral form of Eq.~(\ref{eq:eec0p}) or via mixed jets c.f.  Eq.~(\ref{eq:eecmix}). 
The angular integral of the above correlation vanishes by construction. In Fig.~\ref{fig:eecratio} (upper plot), we show the above correlations, Eq. \eqref{eq:eecc} as functions of $R_L=\theta$ for the typical jet energies at the LHC. It  shows a characteristic behavior in different angular regions. In particular, in the large $\theta$ region toward the jet boundary, the correlation $\langle {\rm EEC} (\theta)\rangle_C$ decreases with $\theta$. This is because both terms in Eq.~(\ref{eq:eecc}) vanish and their difference vanishes too. To expose a possibly interesting correlation signal in this region, it is advantageous to look at the the ratio of the two distributions, 
\begin{equation}
    {\cal R}_{\rm EEC}(\theta)=\frac{d
\langle {\rm EEC}(\theta)\rangle_{\rm full} 
}{d^2{\theta}}/\frac{d \langle {\rm EEC}(\theta)\rangle_{\rm Disc.}}{d^2\theta}  \ ,\label{eq:eecc0}
\end{equation}
where the numerator and denominator come from Eq.~(\ref{eq:eecdef}) and (\ref{eq:eec0p}), respectively. As shown in Fig.~\ref{fig:eecratio} (lower plot), this ratio demonstrates the strength of the correlation at both small and large angle of $\theta$.

\begin{figure}[htbp]
  \includegraphics[scale=0.35]{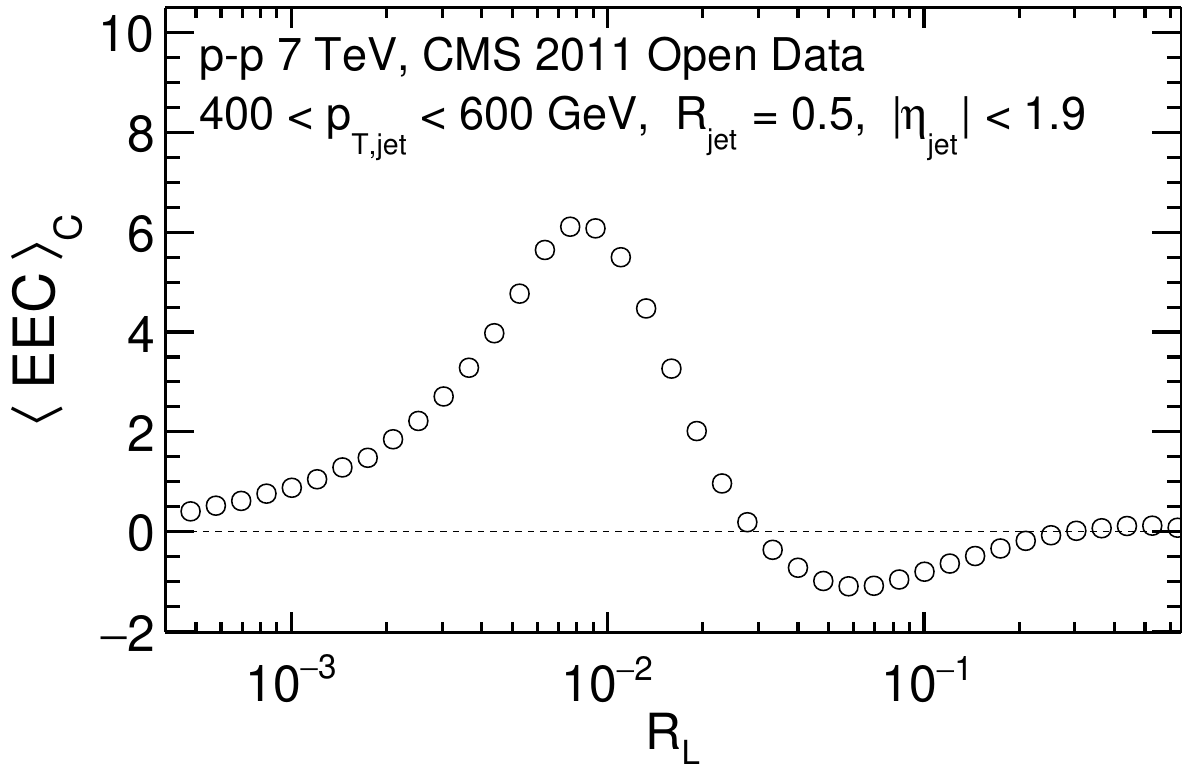} 
   \includegraphics[scale=0.35]{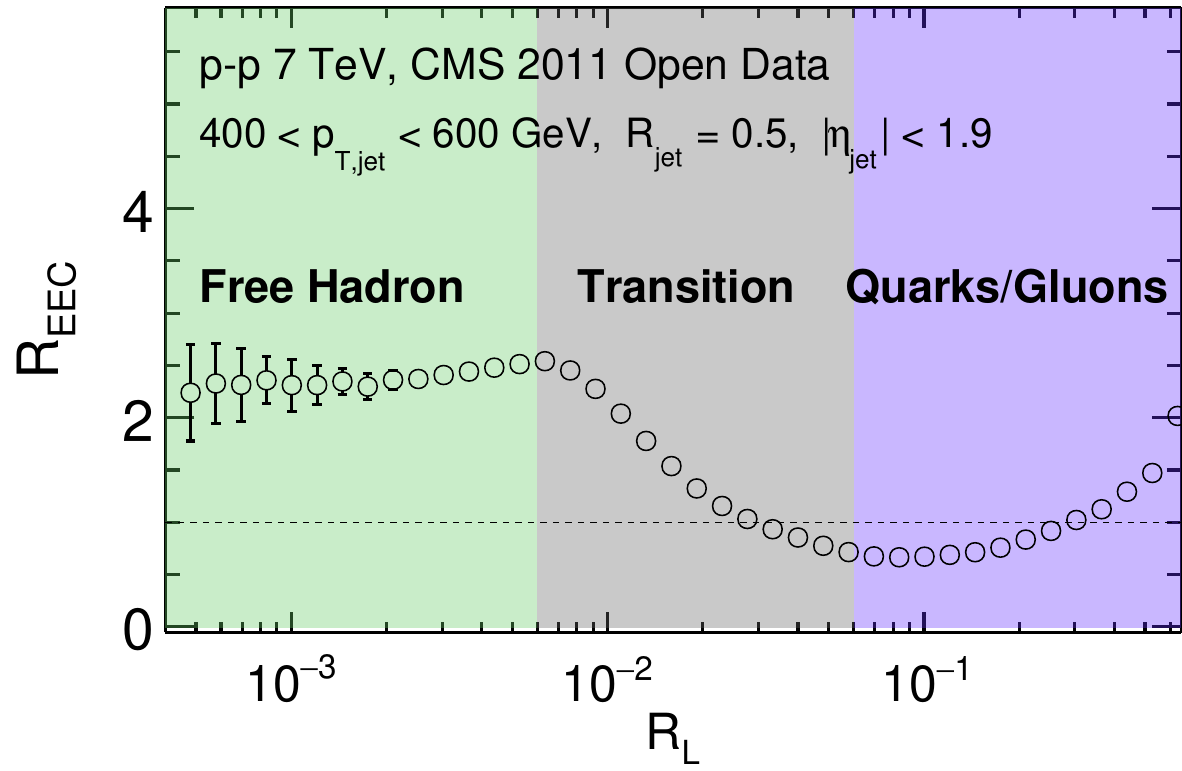} 
\caption{Similar to Figure \ref{fig:eecratio}, but analyzed using the CMS 2011 Open Data~\cite{CMS2011:Open_data}. Only statistical uncertainties are shown.
}
  \label{fig:data}
  \vspace{-3.ex}
\end{figure}

Experimentally, all of the quantities that define the correlations in Eqs.~(\ref{eq:eecc},\ref{eq:eecc0}) can be measured. 
Therefore, these correlations can be systematically studied in experiments at the LHC and RHIC. There are a number of interesting observations from Fig.~\ref{fig:eecratio}:
\begin{itemize}
    \item Stronger correlations at moderate angles for all jet energies can be seen from the ratio curves and
    higher jet energies tend to have larger ratios as well.
    Meanwhile, the genuine correlation increases with jet energy at small angles. This can be seen from both $\langle {\rm EEC}(\theta)\rangle_C$ and ${\cal R}_{\rm EEC}(\theta)$ curves in the figure.
    \item Between these two regions, the lower the jet energies, the ratios are closer to 1. This means that the experiment measured EECs of Eq.~(\ref{eq:eecdef}) is close to the disconnected EECs constrcuted from the individual energy flows Eq.~(\ref{eq:eec0p}).
    \item Three characteristic regions in these plot are very similar to those classified in Ref.~\cite{Komiske:2022enw}, especially for high energy jets: at small angles (the ``Free Hadrons Region"), $R_{\rm EEC}$ is essentially flat, indicating unmodified hadron emissions with negligible collinear effects; at moderate angles (the ``Transition Region"), $R_{\rm EEC}$ decreases, signalling the onset of collinear dynamics and hadronization processes that suppress two‐particle correlations; and at large angles (the ``Collinear Region"), $R_{\rm EEC}$ rises again, driven by perturbative‐QCD collinear splittings that enhance correlations. 
This demonstrates that the above ratio provides an alternative method to study the associated physics.
\end{itemize}

The ratio clearly highlights different physics in different kinematic regions. Experimental measurements of these genuine correlations will provide crucial information on the QCD dynamics involved in the particle correlations in high-energy jets and help to investigate hadronization effects at colliders energies. Especially, when the two hadrons are close to each other, they tend to be produced from a single non-perturbative fragmentation as suggested in Refs.~\cite{Liu:2024lxy,Barata:2024wsu,Chang:2025kgq,Lee:2025okn,Guo:2025zwb}. The results shown in Fig.~\ref{fig:eecratio} expose these  effects very clearly. The same discussion applies to low $p_T$ jets at RHIC energies.

We further analyze the CMS 2011 Open Data sample~\cite{CMS2011:Open_data} and compute the same EEC-based quantities defined in Eqs.(\ref{eq:eecc}) and(\ref{eq:eecc0}). 

In this analysis, we use all particles inside the reconstructed jets, including photons, to ensure that the single-particle energy distribution defined in Eq.~(\ref{eq:ecdef}) is properly normalized (i.e., its integral is unity). The input particles are the particle-flow (PFCs) candidates provided in the CMS 2011 Open Data, which combine information from the tracker, calorimeters, and muon system to reconstruct individual charged and neutral particles. Jets are reconstructed from these candidates with the anti-$k_T$ algorithm with radius parameter $R_{\rm jet}=0.5$. Jets are required to satisfy $|\eta_{\rm jet}|<1.9$, and $400<p_{\rm T,jet}<600~\mathrm{GeV}$, a window chosen to ensure sufficient statistics while restricting to a regime where the observable is theoretically well controlled.
The energy correlator is computed from the PFC candidates within each selected jet.  The correlator is binned in $R_L\in[10^{-4},1.0]$ with 50 logarithmically spaced bins. The disconnected baseline and the genuine correlation are constructed from the same event sample, so that detector effects common to both enter consistently and cancel in their ratio. Figure~\ref{fig:data} shows qualitative agreement between the data and our PYTHIA8 simulations, demonstrating that the proposed measurement is feasible in a realistic hadron-collider environment. In particular, $\langle \mathrm{EEC}\rangle_{C}$ is positive at small angles, becomes negative at intermediate angles, and approaches zero at large angles. The ratio $R_{\rm EEC}$ cleanly separates three regimes, characterized by $dR_{\rm EEC}/dR_L \simeq 0$ (small angles), $dR_{\rm EEC}/dR_L<0$ (intermediate angles), and $dR_{\rm EEC}/dR_L>0$ (large angles), consistent with Ref.~\cite{Komiske:2022enw}.  Comparing Fig.~\ref{fig:data} and~\ref{fig:eecratio}, we find that the transition values of $R_L$ differ between the CMS Open data and the \textsc{Pythia} simulations. We believe this is due to different methods in treating particle decays and the limitation of parton shower in describing small-angle intra-jet correlator structure. 

We also note that the Open Data sample contains only $\mathcal{O}(10^6)$ jets; modern LHC datasets enable substantially higher-precision and more differential measurements. We would like to emphasize that this is an analysis of the public CMS 2011 Open Data and is not an official CMS result. The data have not been corrected for detector effects, and accordingly we refer to this work as an ``analysis" rather than a ``measurement." Our aim is to provide a proof of principle that the proposed observable is robustly extractable from real collider data. We note, moreover, that because our principal results are presented as ratios of self-normalized distributions, a substantial part of the detector-related effects largely cancel between numerator and denominator. This suppresses the sensitivity to detector systematics and supports the robustness of the qualitative features we report.

The correlation measurement proposed above will be very useful in deciphering the underlying mechanism for the nuclear modification of the EEC observables, because it cancels out the single-particle nuclear effects inside the jet. 
For example, we can construct the difference/modification between the correlations of Eq.~(\ref{eq:eecc}) for $AA$ and $pp$ collisions,
\begin{equation}
\Delta{\langle\rm EEC\rangle}^{AA}(\theta)={\langle {\rm EEC}(\theta)\rangle_C^{AA}}-{\langle {\rm EEC}(\theta)\rangle_C^{pp}}\ ,\label{eq:eecaa}
\end{equation}
or a double ratio,
\begin{equation}
   I\!\! {\cal R}_{\rm EEC}^{AA}(\theta)=\frac{\frac{d
\langle {\rm EEC}(\theta)\rangle^{AA}_{\rm full}
}{d^2{\theta}}/\frac{d \langle {\rm EEC}(\theta)\rangle_{\rm Disc.}^{AA}}{d^2\theta}}{\frac{d
\langle {\rm EEC}(\theta)\rangle ^{pp}_{\rm full}
}{d^2{\theta}}/\frac{d \langle{\rm EEC}(\theta)\rangle_{\rm Disc.}^{pp}}{d^2\theta}}  \ .\label{eq:eecraa}
\end{equation}
We expect that the nuclear modification factors will be more evident in the regions of small and large $\theta$ where genuine correlations  are expected to dominate. 

For example, at very small angles, the dominant contribution comes from the jet energy loss. By studying the above double ratio, we may be able to directly extract the fraction of the jet energy loss in these measurements. This is because the correlation ratio is (almost) proportional to the jet energy (see bottom panel of Fig.~\ref{fig:eecratio}) . As a consequence, the ratio of Eq.~(\ref{eq:eecraa}) will be bigger than 1. The jet energy dependence of this double ratio shall provide crucial information on the jet energy loss as a function of the jet energy. Similarly, at large angle, we will be able to study the medium modification of the parton splitting. Again, the reason behind this is that in this region it is dominated by the splitting contributions. 
We have also verified this result using the state-of-the-art linear Boltzmann transport (LBT) model~\cite{Xing:2024yrb}.  Moreover, this method can be straightforwardly extended to small collision systems to search for potential in‐medium effects, free from the systematic uncertainties associated with Glauber‐model scaling. More detailed discussion about the results in heavy-ion collision and the small systems will be presented in the forthcoming follow-up paper.

In addition, it has been realized that the medium responses will modify the EEC of Eq.~(\ref{eq:eecdef}) in heavy-ion collisions~\cite{Bossi:2024qho,Barata:2025fzd}, which might be mis-interpreted as the medium modification of parton splitting in the medium. However, by measuring the genuine correlations, the contribution from the medium response will be canceled out in Eqs.~(\ref{eq:eecc}) and (\ref{eq:eecc0}) in $AA$ collisions, because the medium response has no particular angular dependence, see, for example, the discussions in Ref.~\cite{Barata:2025fzd}. 
That means there is no modification to Eqs.~(\ref{eq:eecaa}) and (\ref{eq:eecraa}). This demonstrates that we can disentangle different physics in the measurements of the EEC in heavy-ion collisions by focusing on the genuine correlations. 

We note that the above discussion applies to the EEC measurements in proton-nuclei collisions as well, where we can define similar correlations. From that, we can study the difference (or ratio) between the correlations in $pA$ collisions and those in $pp$ collisions to investigate the underlying physics behind the nuclear modification observed in experiments~\cite{ALICEpA}.

\textbf{\textit{Conclusion.}} In this paper, we have proposed to study the particle correlations inside high energy jets by constructing a disconnected EEC through the two individual energy flows and comparing that with the measured EEC, as described by Eqs.~(\ref{eq:eecdef}) and (\ref{eq:eec0p}). We have discussed the characteristic features of the ratios/differences defined from them as functions of angle between the two particles. The energy evolution of this ratio will help us to understand the underlying physics behind the energy-energy correlators and may reveal the hadronization effects in high energy jets at colliders. 

While our baseline study uses PYTHIA8 to establish qualitative expectations, the method is not intended as a generator-only proposal. As a proof of principle, we have also analyzed CMS 2011 Open Data and found that the main features of the observable are reproduced in real $pp$ data within the available statistical precision. The robustness of the qualitative features is supported by their appearance in real data.

We further propose to measure the double ratios of Eq.~(\ref{eq:eecraa}) between heavy-ion collisions and proton-proton collisions as a crucial method to disentangle different contributions to the EEC observable. In particular, the small angle behavior of this double ratio will inform the jet energy loss fraction in heavy-ion collisions, and large angle double ratios will tell us about the medium modification of the parton splitting in the medium. On the other hand, we argue that the medium response will mostly be canceled out in these double ratios.

Our proposal could be straightforwardly extended to multi-point energy correlators. A comprehensive investigation of all of these correlators will deepen our understanding of the QCD dynamics involved in these measurements.  Finally, a dedicated study in $e^{+}e^{-}$ collisions, including comparisons across center-of-mass energies and to existing LEP measurements, would provide a particularly clean benchmark of the method. In this Letter we have focused on $pp$/$pA$/$AA$ collisions, motivated by the immediate experimental accessibility at the LHC and RHIC and by the opportunity to quantify cold- and hot-nuclear-matter effects in nuclear collisions; our Open Data analysis demonstrates that the proposed observable can be extracted from real $pp$ data, supporting the feasibility of these dedicated measurements~\cite{SM}.

\begin{acknowledgments}

\textbf{\textit{Acknowledgments}} 
We thank Wenqin Fan, Barbara Jacak, Mateusz Ploskon, and Wenjing Xing for helpful discussions.
This work is supported by the Office of Science of the U.S. Department of Energy under Contract No. DE-AC02-05CH11231. W.B.Z is also supported by the National Science Foundation under grant number ACI-2004571 within the framework of the XSCAPE project of the JETSCAPE collaboration and by the U.S. Department of Energy, Office of Science, Office of Nuclear Physics, within the framework of the Saturated Glue (SURGE) Topical Theory Collaboration.


\end{acknowledgments}

\appendix
\section{``Mixing" Events to Construct the Disconnected Correlations}
\label{sec:appendix_mix}

\begin{figure}[htbp]
  \begin{center}
   \includegraphics[scale=0.45]{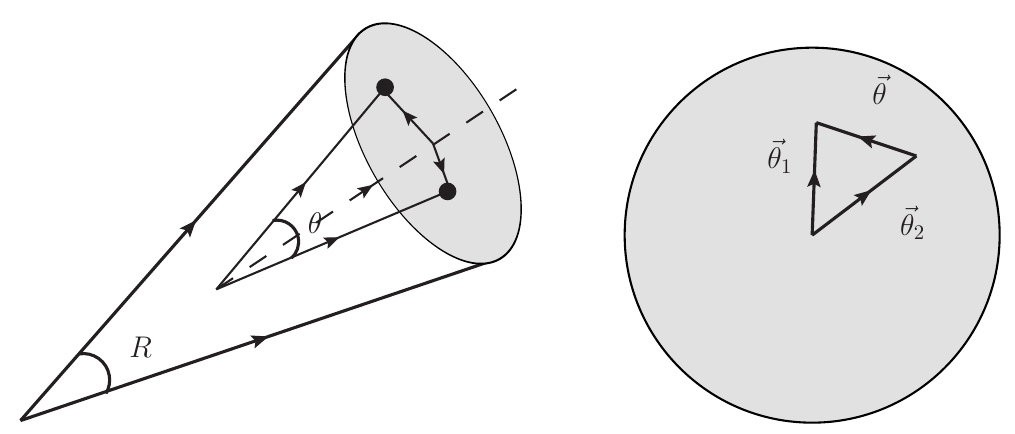} 
\caption{Energy-Energy Correlator measurement inside the jet. The two particles are at relative angles respect to the jet axis: $\vec{\theta}_i$ and $\vec{\theta}_j$. Their angular difference $\vec{\theta}$ is defined as a 2-dimension difference in the plane perpendicular to the jet axis.    }
  \label{fig:Jet_EEC}
 \end{center}
  \vspace{-5.ex}
\end{figure}

As shown in Fig.~\ref{fig:Jet_EEC}, the experimental measurement of the energy-energy correlators is carried out for two particles inside a jet. In the following, we briefly describe the method of event or rather jet mixing used in our work to evaluate the disconnected un-correlated contribution. We use the following procedure: 
\begin{enumerate}
    \item Reconstruct the jets from Pythia and select the jets within certain $P_T$ range, for example $300 {\rm \, GeV} < P_{T,jet} < 350 {\rm \, GeV}$.    
    \item To generate a mixed jet corresponding to a given true jet with $n$ particles, we pick $n$ particles, one each, from $n$ random jets in the same sample, i.e. $P_T$ range, see Fig.~\ref{fig:mixing}. Keep track of the energy fraction carried by each particle and the relative angle $\vec\theta$ respect to the original jet axis,, i.e., $z_i=E_i/E_{J,i}$ and $\vec\theta_i$ for particle $i$ from Jet$_{i}$. 
\begin{figure}[htbp]
  \begin{center}
   \includegraphics[scale=0.65]{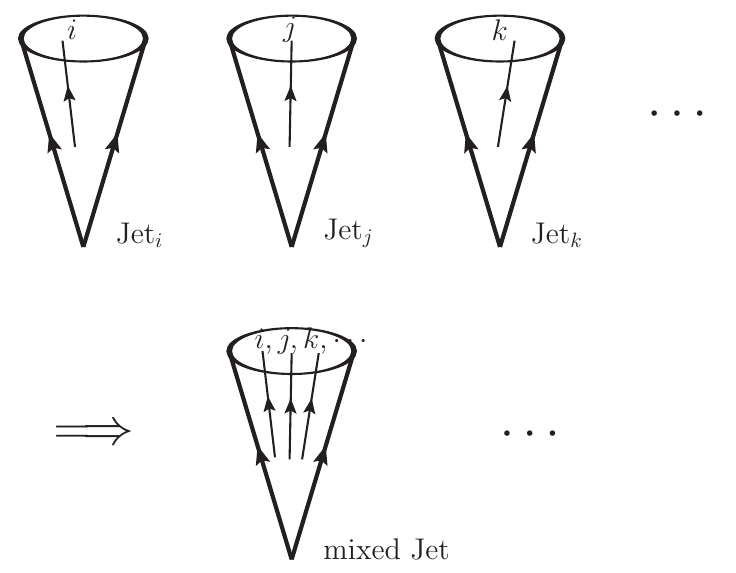} 
\caption{Construct the mixed jet to evaluate the disconnected correlations for the EEC: choosing $n$ particles from randomly chosen jets, such as $i$, $j$, $k$, $\cdots$ from Jet$_i$, Jet$_j$, Jet$_k$, $\cdots$, respectively, to form a mixed jet. We keep the original energy fraction of each particle and their relative angle $\vec\theta$ respect to the original jet axis, e.g., $z_i=E_i/E_{J,i}$ and $\vec\theta_i$ for particle $i$ from Jet$_i$. 
}
  \label{fig:mixing}
 \end{center}
  \vspace{-5.ex}
\end{figure}
    
    \item This procedure will be repeated to have a sufficiently large sample of mixed jets which has the same particle multiplicity distribution within the jets as that of the true jet sample.
    \item Calculate the EEC and the single energy flow based on sample of mixed jets by replacing the true jet sample with that of the mixed events in Eqs. (1) and (6), respectively. For example:

    \begin{eqnarray}
&&    \frac{d\langle {\rm E}(\theta)\rangle_{\rm mix} }{d^2{\theta}}
    =\frac{1}{N_{jet}}\sum_{\rm mixed\, jets\, J }\;\sum_{i\in {\rm J} } \frac{E_{i}}{E_{J,i}} \delta^{(2)}({\vec{\theta}}-\vec{\theta}_i) \nn , \\
&&  \frac{d\langle {\rm EEC}(\theta)\rangle_{\rm mix} }{d^2{\theta}}=\frac{1}{N_{jet}} \sum_{\rm mixed\, jets\, J }\sum_{i\neq j\in {\rm J}} \frac{E_{i}E_{j}}{E_{J,i}E_{J,j}}\nn\\
    &&  ~~~~~~~~~~~~~~~~~~~~~  \times  \delta^{(2)}({\vec{\theta}}-(\vec{\theta}_i-\vec{\theta}_j))\ ,
\end{eqnarray}
for the single energy flow angular distribution and two particle disconnected correlation, respectively. 
\end{enumerate}

\begin{figure}[htbp]
  \begin{center}
   \includegraphics[scale=0.35]{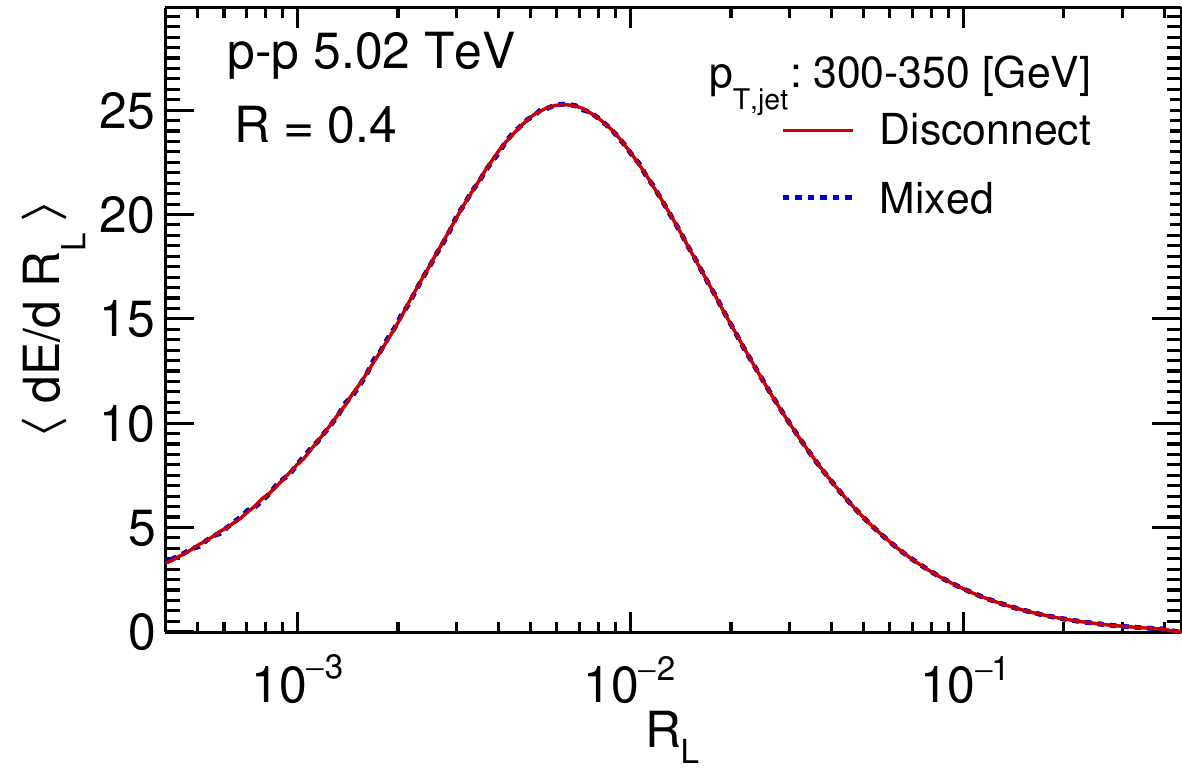}
   \includegraphics[scale=0.35]{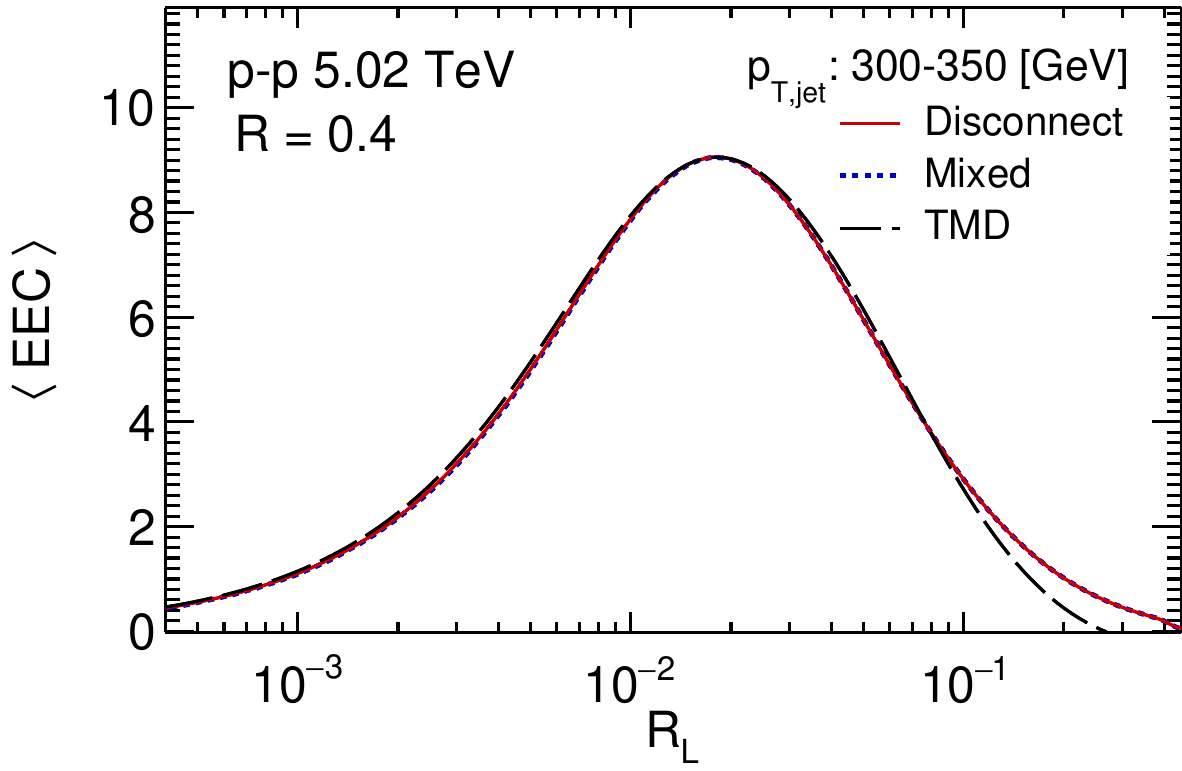} 
\caption{Mixing jet events to evaluate the disconnected correlations: (upper) the angular distribution for single energy flow in the jet for a typical jet energy $P_T=300$-350GeV. (lower) for the disconnected correlations in the same $P_T$ range, comparing the mixing jet events simulation Eq. (8)  with that from integral of single energy flow Eq. (7) in the manuscript and the TMD model calculation (\ref{eq:eecjet}), see below.  }
  \label{fig:eec_mixing}
 \end{center}
  \vspace{-5.ex}
\end{figure}

As shown in Fig.~\ref{fig:eec_mixing} (upper panel), the single energy flow from mixed events is identical to the one obtained from the true jet sample, as it should. Furthermore, as shown in the lower panel of Fig.~\ref{fig:eec_mixing} the EEC from mixed events agree with that obtained from the product of the single energy flows, Eq.~(7) in the manuscript, demonstrating that the EEC from mixed events indeed contains no correlations. As mentioned in the main text, a normalization factor has been included for the jet mixing simulation to ensure the integration results match the value given by Eq. (1) in the manuscript.

\section{TMD Description of Particle Distributions in Jet}

We start with the single energy flow distribution in a jet, i.e., Eq.~(6) in the manuscript.
As shown in Fig.~\ref{fig:Jet_EEC}, the particle inside the jet carries the dominant momentum along the jet direction, where the physics can be described by the associated parton fragmentation functions~\cite{Kaufmann:2015hma,Chien:2015ctp,Neill:2016vbi,Kang:2017glf,Kang:2019ahe,Gao:2024dbv}. In addition, the momentum projection perpendicular to the jet direction can be regarded as a small transverse momentum, $\vec{P}^{h}_{i\perp}=z_iE_J\vec{\theta}_{i}$, where $\theta_i$ represents the angular of hadron $i$ with respect to the jet axis, and $z_{i}$ is its momentum fraction of the jet energy. A small jet radius implies that the transverse momentum of the hadron $|\vec{P}^h_{i\perp}|$ is much smaller than its longitudinal momentum $E_{i}$. Therefore, these distributions can be formulated in terms of the transverse momentum dependent fragmentation functions~\cite{Collins:1981uk,Neill:2016vbi,Kang:2017glf,Kang:2019ahe}.  
Applying the TMD fragmentation functions, we have
\begin{eqnarray}
        \frac{d\langle {\rm E}(\theta)\rangle}{d^2\theta} &=&{\cal N}_1c_1^2E^2\sum_h \int dz_h z_h D_{q/g}^h(z_h,P_{h\perp};\zeta_E) \ , \nn\\ \label{eq:tmd}
\end{eqnarray}
for Eq.~(6) in the manuscript, where $E$ represents the jet energy and $D_{q/g}^{h}(z_h,P_{h\perp})$ with
$P_{h\perp}=c_1z_hE\theta$ denotes the TMD fragmentation function of hadron $h$ within the quark/gluon jet. 
The factors $z_h$ in Eq.~(\ref{eq:tmd}) corresponds to the energy weight in 
Eq.~(6) in the manuscript.
In the  above equation, we introduce a parameter $c_1$ to represent the average energy fraction that the fragmenting quark/gluon carries respect to the jet energy. 
We have also introduced a normalization factor ${\cal N}_1$, which accounts for the contribution from the perturbative matching and has a mild dependence on the jet energy~\cite{Kang:2017glf,Neill:2016vbi,Kang:2019ahe}. 
In addition, the TMD fragmentation function will introduce a dependence on the jet energy represented by the Collins parameter $\zeta_E\sim c_2E$~\cite{Collins:1981uk,Kang:2017glf}, 
\begin{eqnarray}
  D_{q/g}^h(z_h,P_{h\perp};\zeta_E)&=&\int\frac{d^2b_T}{(2\pi)^2}e^{i\frac{P_{h\perp}\cdot b_T}{z_h}}D_{q/g}^h(z_h,\mu_b)\nn\\
  &&\times e^{-\frac{S_{q/g}(c_2E,b_T)}{2}} \ ,
\end{eqnarray}
where $S_{q/g}(\zeta_E,b_T)$ is the associated Sudakov form factor (defined below in Eq.~(\ref{eq:sudakov})) and $D_{q/g}^h(z_h,\mu_b)$ is the integrated fragmentation function at the scale $\mu_b=2e^{-\gamma_E}/b_T$ with $\gamma_E$ the Euler constant. Taking into account the identity $\int dz_h z_h D_{q/g}^h(z_h,\mu)\equiv 1$, we derive the individual energy flow distribution as,
\begin{eqnarray}
\frac{d{\langle {\rm E}(\theta)\rangle}}{d^2\theta}&=&{\cal N}_1c_1^2E^2 \int \frac{d^2b_T}{(2\pi)^2} e^{ic_1E\vec{\theta}\cdot \vec b_T} e^{-\frac{S_{q,g}(c_2E,b_T)}{2}} \ .\nn\\
\label{eq:etheta}    
\end{eqnarray}
In the above equations, the Sudakov form factors take into account all order soft gluon resummation,
\begin{equation}
    S_{q/g}(Q,b)=S_{NP}(b)+\int_{\mu_b^2}^{Q^2}\frac{d\mu^2}{\mu^2}\left[A_{q/g}\ln\frac{Q^2}{\mu^2}+B_{q/g}\right] \ ,\label{eq:sudakov}
\end{equation}
where functions $A$, $B$ can be expanded in terms of $\alpha_s$: $A(\alpha_s) = \sum_{n=1} \left( \frac{\alpha_s}{\pi} \right)^n A^{(n)}$, $B(\alpha_s) = \sum_{n=1} \left( \frac{\alpha_s}{\pi} \right)^n B^{(n)}$. To illustrate the resummation effects, our numerical  calculations take the simple leading terms of $A^{(1)}$ and $B^{(1)}$,
\begin{eqnarray}
A_q^{(1)}&=& C_F, ~~~~A_g^{(1)}=C_A \ , \\
B_q^{(1)}&=& -\frac{3}{2} C_F, ~~~~B_g^{(1)}=-2\beta_0\ .
\end{eqnarray}
In addition, we follow Ref.~\cite{Sun:2014dqm} to parameterize the non-perturbative TMD as
\bea 
S_{NP}(b) = 
\frac{2g_1}{c_1^2} b^2
+ 
{g_2}\ln\left(\frac{b}{b_\ast}\right)
\ln\frac{Q}{\mu_0} \ ,
\eea 
where we have taken the average parton momentum fraction $c_1$ to modify the TMD fragmentation function parameterization and $b_\ast = b/\sqrt{1+b^2/b_{\rm max}^2}$.  For the quark EEC, we follow Ref.~\cite{Sun:2014dqm} to set $g_1 = 0.042{\rm GeV}^2$, $g_2 = 0.84$, $\mu_0 = 1.55 \>{\rm GeV}$, and $b_{\rm max} = 1.5\> {\rm GeV}^{-1}$. For the gluon counterpart, $g_1$ and $g_2$ will be multiplied by a factor $C_A/C_F$.

Finally, substituting Eqs.~(\ref{eq:etheta}) into Eq.~(7) in the manuscript, we will be able to derive the disconnected EEC in terms of the TMD fragmenation functions,
\begin{eqnarray}
  &&   \frac{d}{d^2\theta} \langle {\rm EEC}(\theta)\rangle_{\rm Disc.}\nn\\
     &&=
     {\widetilde{\cal N}c_1^2E^2}\int db b J_0(c_1Eb\theta)e^{-S_{q,g}(c_2 E,b)} 
     \ , \label{eq:eecjet}
\end{eqnarray}
where $\widetilde{\cal N}=N_{\rm Disc.}{\cal N}_1^2$. Again, this parameter will have a mild energy dependence to account for the perturbative matching to the TMD fragmentation functions in the jet evolution.

\begin{figure}[htbp]
\vskip 0.5cm
   \includegraphics[scale=0.4]{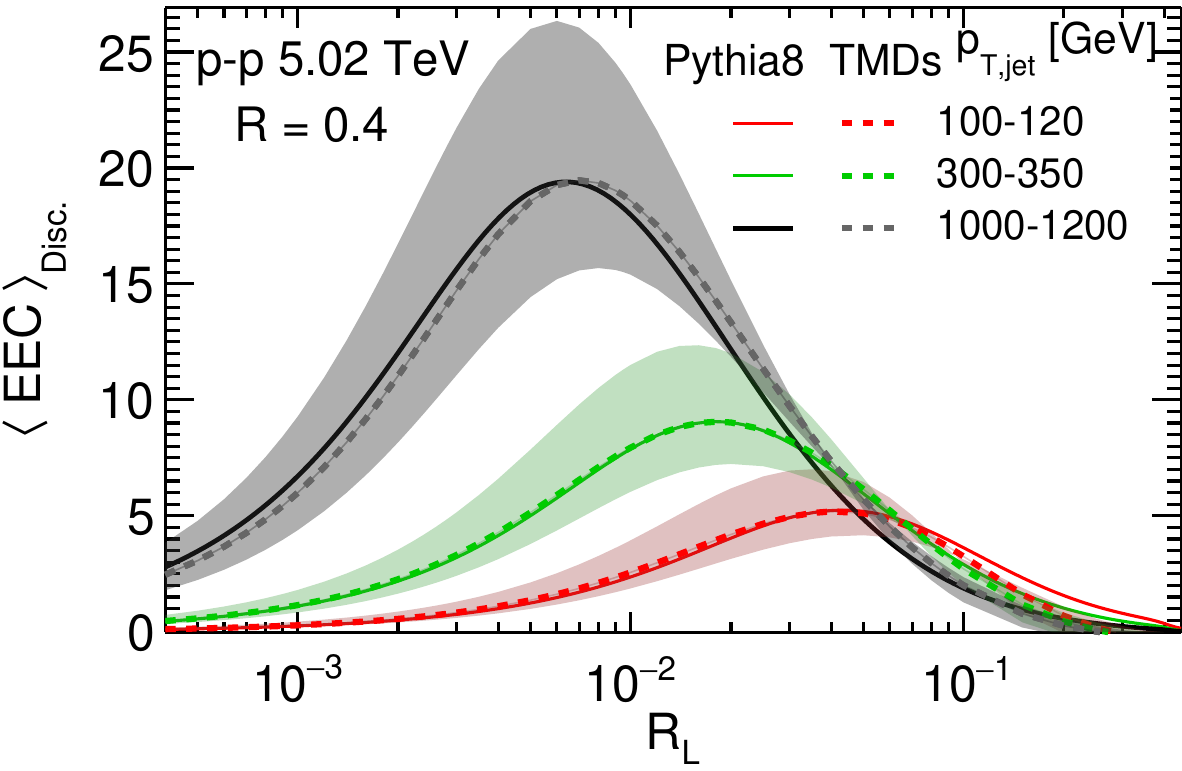} 
\caption{Comparisons between PYTHIA8 simulations and TMD calculations for the `Disconnect' energy correlator. Here the theory bands are obtained by varying the ``scale" parameter $c_2$ in Eq.~(\ref{eq:eecjet}) around the central value of $c_2=0.3$: $c_2\sim 0.20-0.40$. 
}
  \label{fig:EECTMDs}
  \vspace{-3.ex}
\end{figure}

As an illustration, we adopt the above simple model application of the TMD fragmentation contribution and fit the parameters to the PYTHIA simulations for a wide range of jet energies at the LHC: $P_{T,\rm jet}=100-120\rm GeV$, $300-350\rm GeV$ and $1000-1200\rm GeV$, see, Fig.~\ref{fig:EECTMDs}. In the above model, there are three parameters: $C_1$ describes the average energy fraction of the jet carried by the fragmenting parton, $C_2$ represents the hard momentum scale for the Sudakov resummation and an overall normalization ${\cal N}$. All these parameters are expected to be order $1$. In the fit, we find that all three jet energies EECs can be described by the same $C_1=0.62$ and $C_2=0.30$ but with varying normalization factor ${\cal N}$: 0.53 for $E_T=100$, $0.65$ for $E_T=300$ and 0.79 for $E_T=1000$. Clearly, the TMD model calculations capture the main behavior of the disconnected correlators at small angles. 

In addition, we can estimate the uncertainties by varying the fitting parameters. Among these parameters, $C_2$ represents the factorization scale dependence in the TMD model calculations of the disconnected correlations. It is appropriate to estimate the theoretical uncertainties due to this scale dependence. In Fig.~\ref{fig:EECTMDs}, we include the uncertainties bands for three typical jet energies at the LHC, by varying $C_2$ around the central value of $C_2=0.30$: $C_2\sim 0.20-0.40$ but keeping other parameters at their central values. We have also checked the sensitivities of our results on the non-perturbative parameter $g_h$ and they show little dependence for the jet energies in our studies.

We also notice that there have been new theoretical developments on the TMD application to the one-point energy correlator in jets~\cite{Mi:2025abd}, where a sophisticated theory framework has been demonstrated, including next-to-leading order corrections. One should be able to apply these results to the disconnected correlations of two-point energy correlators. This will help to reduce the theoretical uncertainties shown in Fig.~\ref{fig:EECTMDs}. We will leave this for a future publication. 

\bibliographystyle{JHEP-2modlong.bst}
\bibliography{refs}

\end{document}